\documentclass[aps,prl,twocolumn,groupedaddress]{revtex4}

\usepackage{graphicx}
\usepackage{amssymb}

\newcommand{\rs}{\rm \scriptscriptstyle}

\begin{document}

\title{Spectroscopy of Superfluid Pairing in Atomic Fermi Gases}

\author{H.P.\ B\"uchler$^{1}$}
\author{P. Zoller$^{1}$}
\author{W. Zwerger$^{2}$}
\affiliation{$^{1}$Institute for Quantum Optics and Quantum Information
of the Austrian Academy of Science, 6020 Innsbruck, Austria \\
$^{2}$Institute for Theoretical Physics, Universit\"at
Innsbruck, 6020 Innsbruck, Austria}

\date{\today}

\begin{abstract}
We study the dynamic structure factor for density and spin within
the crossover from BCS superfluidity of atomic fermions to the
Bose-Einstein condensation of molecules. Both structure factors
are experimentally accessible via Bragg spectroscopy,
and allow for the identification of the pairing mechanism: 
the spin structure factor allows for the determination of the two
particle gap, while the collective sound mode in the density structure
reveals the superfluid state.
\end{abstract}


\maketitle

Atomic fermions have attracted a lot of interest as current
cooling techniques allow for the creation of molecular condensates
\cite{jochim03,greiner03,zwierlein03,bourdel04}.
These superfluids behave very much like standard Bose-Einstein
condensates (BEC):  the condensate  may be inferred from the momentum distribution
measured in a time of flight experiment. The tunability of the
interaction through a Feshbach resonance then offers the possibility
to explore the crossover from  BEC of tightly bound molecules to the BCS superfluid
state, where Cooper pairs only exist due to many body effects
\cite{nozieres85,randeria90,haussmann93,pistolesi94,holland01,ohashi02}. Recent experiments have
entered this regime \cite{regal04,zwierlein04,grimm04}, however, clear signatures for extended
Cooper pairs in a BCS like ground state are missing so far.
In this letter, we present  a generalization of available
spectroscopic tools to measure the dynamic structure factor for
density and spin, which reveal important
information on the pairing mechanism within the BEC-BCS crossover.

In conventional superconductors, the main characteristic
properties are dissipation free transport and the
Meissner-Ochsenfeld effect, which reveal themselves in the current
response; density fluctuations are suppressed due to
long-range Coulomb interactions \cite{schriefferbook} (see
Ref.~\cite{pistolesi94} for the current response in the BEC-BCS
crossover). In contrast, for uncharged atomic gases transport
measurements are not readily accessible due to the trapping
potential. Then, the dynamic structure factors for density and spin
are suitable quantities for the characterization of the superfluid
ground state within the BEC-BCS crossover. Both quantities are
accessible in traps: recent experiments measured the dynamic
structure factor in interacting Bose gases via Bragg spectroscopy
\cite{stamper-kurn99,steinhauer02,stoferle03}, while the dynamic
spin susceptibility may be inferred by measuring the spin flip
rate in stimulated Raman transitions \cite{torma00,grimmPD}. 
In this paper, we analyze the dynamic structure factor $S_{\rs C}$
and the dynamic spin structure factor $S_{\rs S}$ within the
BEC-BCS crossover. We find, that the dynamic spin structure factor 
is dominated by processes which break paired fermions into two single 
particles and therefore reveals the many-body excitation gap. 
Furthermore, it provides the density of states,  which signals the BCS 
pairing mechanism via the the appearance of a van Hove singularity.
The observation of this singularity was a fundamental indication for
the validity of  BCS theory in conventional superconductors
\cite{schriefferbook}. In turn, the superfluid transition is
characterized by the appearance of a collective sound mode in the dynamic structure
factor; this collective mode has recently been studied
\cite{minguzzi01,ohashi02,hofstetter02}.

An interacting atomic gas of fermions with density  $n_{\rs
F}=k_{\rs F}^{3}/3 \pi^2$ and two different spin states is
characterized by the scattering length $a_{\rs F}$ allowing to
tune the BEC-BCS crossover via the dimensionless parameter
$1/(k_{\rs F}a_{\rs F})$. As shown by Nozi\`eres and Schmitt-Rink
\cite{nozieres85}, the BCS wave function becomes exact
in the BCS limit ($1/(k_{\rs F}a_{\rs F})\ll -1$) and the BEC
limit ($1/(k_{\rs F}a_{\rs F}) \gg 1$). In the following, we use
this BCS wave function to determine the dynamic structure factor
$S_{\rs C}$ from the density response function $\chi_{\rs C}$, and
the dynamic spin structure factor $S_{\rs S}$ from the spin
susceptibility $\chi_{\rs S}$. This wave function accounts for the
pairing between two fermions, while the residual
interaction between unbound fermions providing particle-hole
scattering is neglected. The interaction between molecules (Cooper
pairs) is accounted for within Born approximation
\cite{randeria90} (it is not necessary to introduce a molecular
field as done in Ref.~\cite{ohashi02}). Within the BCS variational
wave function, the fermionic normal Green's function $\mathcal{G}$
and the anomalous Green's function $\mathcal{F}^{+}=-\mathcal{F}$ take the standard
form \cite{schriefferbook}
\begin{eqnarray}
 \mathcal{G}(\Omega_{s},{\bf k}) \!= \!\frac{i \Omega_{s}\!+ \!\epsilon_{
 \bf k}}{(i \Omega_{s})^2\!- \!E_{\bf k}^{2}}, \hspace{15pt}
  \mathcal{F}^{+}(\Omega_{s},{\bf k})\! =\! \frac{- i 
  \Delta}{(i\Omega_{s})^2\!-\!E_{\bf k}^{2}}. \nonumber
\end{eqnarray}
%
%
%
Here, $E_{\bf k}= \sqrt{\epsilon_{\bf k}^2+\Delta^2}$ denotes the
single-particle excitation energy with $\epsilon_{\bf k}= \hbar^2
{\bf k}^2/2m- \mu$  the free fermionic dispersion relation, while
$\Omega_{s}= \pi T (2 s+1) $ denotes fermionic Matsubara
frequencies. The presence of a condensate and superfluid response
in the ground state is encoded in a finite BCS gap $\Delta$. The
gap $\Delta$ is determined by the scattering length $a_{\rs F}$
via the gap equation, and the chemical potential is fixed by the
particle density $n_{\rs F}$ \cite{randeria90},
\begin{eqnarray}
   -1&=&\frac{4 \pi \hbar^2 a_{\rs F}}{m} \int\frac{d{\bf
   q}}{(2\pi)^3}\left[\frac{\tanh\left(\beta E_{\bf q}\right/2)}{2 E_{\bf
   q}}- \frac{m}{\hbar^2 {\bf q}^2}\right] , \label{gapequation}\\
 n_{\rs F} &=& 2 \int\frac{d{\bf q}}{(2
 \pi)^3}\left[\frac{1}{1+e^{\beta E_{\bf q}}}\frac{\epsilon_{\bf
 q}}{E_{\bf q}}+ \frac{E_{\bf q}-\epsilon_{\bf q}}{2 E_{\bf
 q}}\right]
 \label{densityequation} .
\end{eqnarray}
In the BCS limit at low temperatures ($T=0$), these relations give
the chemical potential  $ \mu= \epsilon_{\rs F}$ and the gap
$\Delta = \epsilon_{\rs F} (8 /e^2) \exp\big(\pi/2 k_{\rs F}
a_{\rs F} \big)$ with $\epsilon_{\rs F}$ the Fermi energy. In
turn, in the BEC limit the appearance of a two particle bound
state with binding energy $\epsilon_{0}=\hbar^2/(m a_{\rs F}^2)$
and internal wave function $\phi_{\bf k}=(2 n_{\rs F})^{-1/2}
\Delta/E_{\bf k}$ modifies the chemical potential $\mu=
-\epsilon_{0}/2$ and the gap $\Delta= \epsilon_{0}\sqrt{4\pi
n_{\rs F} a_{\rs F}^3 }$. Within the crossover regime, we solve
Eq.~(\ref{gapequation}) and (\ref{densityequation})
numerically to determine $\Delta(k_{\rs F} a_{\rs F})$ and
$\mu(k_{\rs F}a_{\rs F})$. Note, that in
general the BCS gap $\Delta$ differs from the two particle
excitation gap $\Delta_{\rs S} = \min_{\rs {\bf k}} 2 E_{\bf k}$.

\begin{figure}[hbtp]
\vspace{-0.cm}
\includegraphics[scale=0.35]{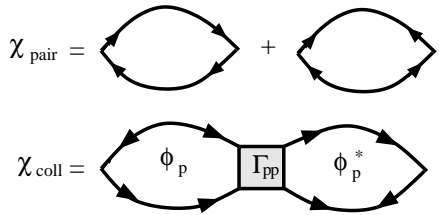}
\vspace{-0.2cm}
  \caption{ \label{diagrams} Diagrams contributing to the density
  response function and the spin susceptibility; $\chi_{\rs pair}$ accounts
  for particle-hole excitations, while the diagrams in $\chi_{\rs coll}$
  describe collective sound excitations.}
\end{figure}

In the following, we calculate the dynamic structure factor $
S_{\rs C}=- {\rm Im}\chi_{\rs C}/\pi $ and the dynamic spin
structure factor $S_{\rs S}=-{\rm Im}\chi_{\rs S}/\pi $ via their
relations to the density response function  and the spin
susceptibility, respectively. The density response $\delta
\rho_{\rs C}(\omega,{\bf k})$  to a small external drive $\delta
V_{\rs C}(\omega,{\bf k})$  follows from $\delta \rho_{\rs
C}(\omega,{\bf k})= \chi_{\rs C}(\omega,{\bf q})\delta V_{\rs
C}(\omega,{\bf k})$
%
%
with the linear response function (in real space)
\begin{equation}
  \chi_{\rs C}(t,{\bf x})= - i \Theta(t)
  \langle \left[ \rho_{\rs C}(t,{\bf x}),\rho_{\rs C}(0,0)\right]\rangle .
\end{equation}
Here, $\langle \ldots \rangle$ denotes the quantum statistical
average at fixed temperature $T$ and chemical potential $\mu$,
while the density operator is defined by $\rho_{\rs C} =
\psi^{+}_{\uparrow} \psi^{}_{\uparrow}+\psi^{+}_{\downarrow}
\psi^{}_{\downarrow}$. The analogous definition applies for the
spin susceptibility with the density $\rho{\rs C}$
replaced by the spin density $\rho_{\rs S}= \psi^{+}_{\uparrow}
\psi^{}_{\uparrow}-\psi^{+}_{\downarrow} \psi^{}_{\downarrow}$.
The diagrams contributing to the response functions $\chi_{\rs C}$
and $\chi_{\rs S}$ are shown in Fig.~\ref{diagrams}. 
Note, that
each diagram has to be weighted by a nontrivial factor $\pm 1$.
These factors differ for the response function $\chi_{\rs C}$ and
the spin susceptibility $\chi_{\rs S}$, and provide different
cancellations between diagrams.  We distinguish between two types
of diagrams: The first type of diagrams $\chi_{\rs pair}$ involve
only normal and anomalous Green's function and describe two
particle excitations, while the second type of diagrams $\chi_{\rs
coll}$ involve the vertex operator $\Gamma_{\alpha \beta}$ and
account for collective excitations. The BCS wave function
neglects particle-hole scattering and $\Gamma$ accounts only for
particle-particle (hole-hole) scattering. Then, $\alpha, \beta \in
\{p,\overline{p}\}$ describe the incoming and outgoing type of
particles; $p$ accounts for particles and $\overline{p}$ for
holes. The vertex operator $\Gamma$ has to be calculated with the
help of the BCS  wave function, see below. Note, that
Fig.~\ref{diagrams} shows only the diagram involving $\Gamma_{p
p}$, while the other diagrams exhibit a similar structure. In the
following, we are mainly interested in the zero temperature limit
$T=0$ and in the low momentum regime $k \ll 1/\xi =\Delta/\hbar
c_{s}$ (the energy of the collective modes is below the two
particle gap $\omega < \Delta_{\rs S}$). Here, $c_{s}$ denotes the
macroscopic sound velocity and $\xi$ the size of the pairs with 
$\xi \sim \hbar v_{\rs F}/\Delta$
in the BCS limit and  $\xi \sim a_{\rs F}$ in the BEC limit.
For typical experimental setups, the scale $\xi$ is
small compared to the trap size $R$, and we can safely assume the
condition $1/R<k<1/\xi$. Then, the trapping potential plays a
minor role as has been shown in the measurement of the dynamic
structure factor and the tunnelling probability, see
Ref.~\cite{stamper-kurn99,steinhauer02,grimmPD}. Therefore, we
ignore the influence of a trapping potential in the following
analysis.

\begin{figure}[hbtp]
\vspace{-0.cm}
\includegraphics[scale=0.4]{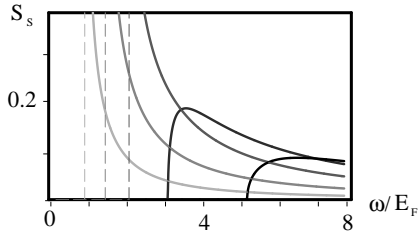}
\vspace{-0.2cm}
  \caption{ \label{pairBCS-BEC}
  Spin structure factor $S_{\rs S}(\omega)$ in units $\hbar n_{\rs F}/E_{\rs
  F}$ for different scattering lengths $1/(a_{\rs F} k_{\rs F})=
  -0.5,-0.25,0,0.25,0.5$. The spin structure is quenched below the
  spin gap $\omega < \Delta_{\rs S}$, and exhibits a characteristic
  singularity in the BCS limit.
 }
\end{figure}

First, we focus on the spin susceptibility  $\chi_{\rs S}$. At
zero momentum, the spin susceptibility $\chi_{\rs S}(\omega,0)$ is
equivalent to the response driven by the spin flip Hamiltonian $H=
\lambda(t) \int d{\bf x} \big[ \psi^{+}_{\uparrow}
\psi^{}_{\downarrow} +\psi^{+}_{\downarrow}
\psi^{}_{\uparrow}\big]$. A perturbation of this form is realized
experimentally by driving a stimulated Raman transition between the
two hyperfine states of the two component Fermi gas \cite{grimmPD}. 
%
Within the diagrammatic expansion of $\chi_{\rs
S}$, the diagrams in $\chi_{\rs coll}$ cancel each other and only
pair excitations $\chi_{\rs pair}$ survive;  their contribution
takes the form
\begin{eqnarray}
 \chi_{\rs S}(\Omega_{s},{\bf k}) \!\!\!&=&\!\! - 2 T \sum_{t\in {\bf Z}}
 \int \frac{d{\bf q}}{(2\pi)^{3}} \Big[\mathcal{G}(\Omega_{t},{\bf q})
 \mathcal{G}(\Omega_{s+t},{\bf q+k})  \nonumber
 \\ && \hspace{40pt} +
 \mathcal{F}^{+}(\Omega_{t},{\bf q}) \mathcal{F}(\Omega_{s+t},{\bf
 q+k}) \Big] \label{spinsusceptibility}
\end{eqnarray}
(Note, that for the density response function $\chi_{\rs C}$ the
$+$ sign between the two terms is replaced by a $-$ sign
\cite{schriefferbook}.) The integration in
(\ref{spinsusceptibility}) involves standard methods, and we
present here only the final result for the spin structure factor
in the low momentum limit $k \ll 1/\xi$, see
Fig.~\ref{pairBCS-BEC},
\begin{equation}
   S_{\rs S}(\omega) \!=\! \frac{3}{4}
   \frac{(n_{\rs F}/\epsilon_{\rs F})\Delta^2}{\omega \sqrt{\left(\hbar \omega/2\right)^2\!-\!\Delta^2}}
   \left[\frac{\sqrt{\left(\hbar \omega/2\right)^2\!-\! \Delta^2}\!+\!
   \mu}{\epsilon_{\rs F}}\right]^{1/2} \hspace{-12pt}.
   \label{spinstructurefactor}
\end{equation}
The spin structure factor exhibits a gap $\Delta_{\rs S}= 2
\min_{\rs {\bf k}} E_{\bf k}$, i.e., for positive $\mu$ the spin gap
$\Delta_{\rs S}= 2 \Delta$ is determined by the BCS gap, while for
$\mu<0$, it is given by the binding energy $\Delta_{\rs S}=
2 \sqrt{\mu^2+\Delta^2}$ approaching $\epsilon_{0}$ in the BEC limit. 
The shape of the spin structure factor differs in
the two limiting regimes: the shape exhibits the characteristic
$1/\sqrt{\omega-2 \Delta}$ singularity of the density of states
for the BCS pairing mechanism, while in the BEC regime $S_{\rs S}$
exhibits a maximum. In the
crossover regime, Eq.~(\ref{spinstructurefactor}) smoothly
interpolates between these two limits. Next, we analyze the
modifications of the spin structure factor for  temperatures above
the superfluid transition temperature $T_{c}$. Then, Eq.~(\ref{gapequation}) 
implies $\Delta=0$, and the spin structure
factor in the BCS limit reduces to that of a Fermi
gas. In turn, the presence of bound fermion
pairs dominate the spin structure factor even above $T_{c}$ ($T<T^{*}$) and $S_{\rs S}$
exhibits a pseudo gap $\sim \Delta_{\rs S} \approx \epsilon_{0}$ with dissociated fermion
pairs providing a small but finite weight within the gap. The
pseudo gap finally disappears above the pairing temperature $T>
T^{*}$ via a smooth crossover. Therefore, the measurement of the
spin structure factor provides a suitable tool for the
characterization of the pairing temperature $T^{*}$ and the spin gap
$\Delta_{\rs S}$.

\begin{figure}[hbtp]
\vspace{-0.cm}
\includegraphics[scale=0.32]{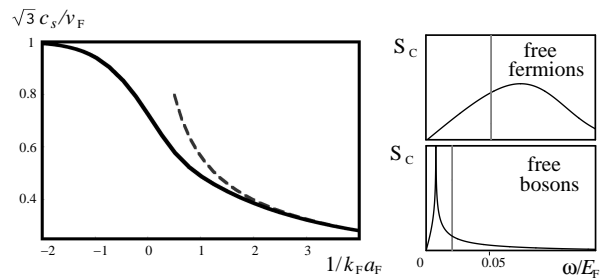}
\vspace{-0.2cm}
  \caption{ \label{soundvelocity} Left: Sound velocity $c_{s}$ in the
  BEC-BCS crossover in units $v_{\rs F}/\sqrt{3}$. The dashed line is
  the Bogoliubov sound velocity. Right: Structure factor $S_{\rs
  C}(\omega)$ of weakly
  interacting fermions and bosons above the superfluid transition
  temperature
  $T=\epsilon_{\rs F}/2>T_{c} $ and $k=\sqrt{3}k_{\rs F}/40$. The gray line
  denotes the peak of the sound mode below the superfluid transition. }
\end{figure}

In contrast, the dynamic density structure factor $S_{\rs C}$ in
superfluids is dominated by a collective sound excitation
representing the Goldstone mode of the broken symmetry (in
conventional superconductors, Coulomb interactions lift this mode to the plasma
frequency). The dynamic structure
factor at low momenta $k \ll 1/\xi$ takes the form
\begin{equation}
  S_{\rs C}(\omega,{\bf k})= n_{\rs F}\frac{\hbar  k}{2 m c_{\rs s}}
  \delta\left(\omega -c_{\rs s} k \right) \label{collectiveexcitations}
\end{equation}
exhausting the $f$-sum rule and compressibility sum rule,
\begin{eqnarray}
 \int_{0}^{\infty} d\omega \hbar \omega S_{\rs C}(\omega, {\bf k})&
=& n_{\rs F} \frac{\hbar^{2} {\bf k}^2}{2
 m}, \label{fsumrule}\\
\lim_{k\rightarrow 0} \int_{0}^{\infty} d\omega \frac{S_{\rs
C}(\omega, {\bf k})}{\hbar \omega} &=& \frac{n_{\rs F}}{2 m
c_{s}^{2}}. \label{compressibilitysumrule}
\end{eqnarray}
The determination of the sound velocity within the BEC-BCS
crossover requires the calculation of the diagrams $\chi_{\rs
coll}$ in Fig.~\ref{diagrams}; its contribution exhausts the sum
rules (\ref{fsumrule}) and (\ref{compressibilitysumrule}) at small
momenta $k \ll 1/\xi$ and frequencies $\hbar \omega < \Delta_{\rs
S}$ i.e., a cancellation appears between $\chi_{\rs pair}$ and
$\chi_{\rs coll}$ at frequencies $\omega > \Delta_{\rs S}$. The
diagrams in $\chi_{\rs coll}$ provide the response function
\begin{equation}
  \chi_{\rs C}(\omega,{\bf k}) = 4 \sum_{\alpha,\beta}
  \phi_{\alpha}^{*}(\omega,{\bf k}) \Gamma_{\alpha,\beta}(\omega, {\bf
  k})\phi_{\beta}(\omega,{\bf k}), \label{densityresponse}
\end{equation}
where one factor 2 accounts for summation of spin indices, while
the second factor 2 appears as  each vertex operator
$\Gamma_{\alpha \beta}$ contributes to  two diagrams. Using the
microscopic approach  by Haussmann \cite{haussmann93},
the vertex operator $\Gamma_{\alpha \beta}$ takes the form
\begin{equation}
 (\Gamma_{\alpha \beta})^{-1}=
  \left[ \frac{m}{4 \pi \hbar^2 a_{\rs F}} - \int \frac{d{\bf
  k}}{(2\pi)^3} \frac{m}{\hbar^2{\bf k}^2}\right] \delta_{\alpha \beta}
  + M_{\alpha \beta}
\end{equation}
with
\begin{eqnarray}
 M_{p p}(\Omega_{s},{\bf k})&\!=\!&  T \sum_{t\in {\rm Z}}\int{\frac{d{\bf q}}{(2\pi)^3}}
 \left[\mathcal{G}(\Omega_{s-t},{\bf
 q\!+\!k})\mathcal{G}(\Omega_{t},-{\bf q})\right] \!,  \nonumber\\
 M_{p \overline{p}}(\Omega_{s},{\bf k})&\!=\!& T \sum_{t\in {\rm Z}}\int{\frac{d{\bf q}}{(2\pi)^3}}
 \left[\mathcal{F}(\Omega_{s-t},{\bf
 q\!+\!k})\mathcal{F}(\Omega_{t},-{\bf q})\right]\!, \nonumber
\end{eqnarray}
with $M_{p p}=M_{\overline{p} \overline{p}}^{*}$ and $M_{p
\overline{p}}=M_{\overline{p}p}^{*}$.
 The propagators $\phi^{*}_{p}$
($\phi_{p}$) account for the creation (destruction) of a fermion
pair from the condensate, and take the form
\begin{equation}
 \phi_{p}=T \sum_{t} \int \frac{d{\bf q}}{(2 \pi)^3}
 \mathcal{G}(\Omega_{t},{\bf q}) \mathcal{F}(\Omega_{t+s},{\bf q+k}) .
\end{equation}
Solving the above equations in the limit of small frequency
$\omega \ll \Delta_{\rs S}$ and momenta $k\ll 1/\xi$ provides the
collective sound mode with  sound velocity $c_{s}$; the result is
shown in
Fig.~\ref{soundvelocity}.

Within the BCS limit, the collective excitations
(\ref{densityresponse}) provide the Bogoliubov-Anderson sound mode for a
neutral superconductor with $c_{s}=v_{\rs F}/\sqrt{3}$
\cite{anderson58}. The particle-hole contributions, dominating the
structure factor for a weakly interacting Fermi gas, are quenched
due to the opening of the excitation gap. Note, that the leading
correction to the sound velocity $c_{s}=v_{\rs F}/\sqrt{3}[1-8
k_{\rs F}a_{\rs F}/\pi]$ derives from  particle-hole scattering
\cite{anderson58}; such scattering processes are not contained in
our approach. In turn, within the BEC limit
Eq.~(\ref{densityresponse}) gives the structure  factor $ S^{\rs
BEC}_{\rs C}(\omega,{\bf k})= 2 n_{\rs F}\delta\big(\omega- \hbar
{\bf k}^2/4m\big)$. The variational BCS wave function approach
provides the zeroth order result describing a noninteracting
Bose gas of molecules with density $n_{\rs B}= n_{\rs F}/2$ and
mass $m_{\rs B}=m/2$. The structure factor  becomes 4
times the structure factor of an ideal Bose gas. This factor 4
appears as the external potential drives the fermionic density
operator instead of the bosonic density operator; the structure
factor satisfies the $f$-sum rule for fermions with $n_{\rs F}/m=
4 n_{\rs B}/m_{\rs B}$. Going beyond leading order, the above
equations also incorporate the repulsion between the bound pairs, 
and provide the structure factor $S^{\rs BEC}_{\rs
C}(\omega,{\bf k})= 2 n_{\rs F} \tilde{\epsilon}_{\bf
k}/\tilde{E}_{\bf k}\delta\big(\omega- \tilde{E}_{\bf
k}/\hbar\big)$ with $\tilde{\epsilon}= \hbar^2 {\bf k}^2/2 m_{\rs
B}$ and the Bogoliubov excitation spectrum $\tilde{E}^2_{\bf
k}=\tilde{\epsilon}_{\bf k}^{2}+ 2 \mu_{\rs B}
\tilde{\epsilon}_{\bf k}$. The structure factor describes a weakly
interacting Bose gas with sound velocity $c_{\rs s}=\sqrt{\mu_{\rs
B}/m_{\rs B}}$. Here, $\mu_{\rs B}= 4 \pi \hbar^2 a_{\rs B} n_{\rs
B}/ m_{\rs B}$ denotes the bosonic chemical potential accounting
for the scattering length $a_{\rs B}= 2 a_{\rs F}$ within Born
approximation; its exact value $a_{\rs B}\approx 0.6$ has recently
been derived \cite{petrov03}.

Next, we focus on the dynamic structure factor above the
superfluid transition temperature  $T>T_{c}$ and compare it with
the  structure factor Eq.~(\ref{collectiveexcitations}). We first
focus on the collisionless regime $\omega \tau
> 1$ with $\tau$ the collision time; this limit is naturally
achieved for weakly interacting atomic gases ($ |k_{\rs F}a_{\rs
F}|< 1$) and frequencies $\omega$ above the trapping frequency
\cite{pitaevskiibook}. Within the BCS limit, the system reduces to
a Fermi liquid with a weak attractive interaction. 
The structure factor exhibits the particle-hole excitation spectrum of a 
weakly interacting Fermi gas at finite temperature (interactions only renormalize the Fermi
velocity $v_{\rs F}$), while the zero sound mode is overdamped for
attractive fermions; the structure factor is shown in
Fig.~\ref{soundvelocity}. In turn, in the BEC limit the system
above the critical temperature $T_{c}$ reduces to a gas of bosonic
molecules. The structure factor of a degenerate Bose gas with
temperatures above the superfluid transition temperature derives
from the bosonic Lindhard function; the structure at low momenta
is shown in Fig.~\ref{soundvelocity}. Comparing these structure
factors with Eq.~(\ref{collectiveexcitations}), we find that the
superfluid state is characterized by the appearance of a
collective sound mode. However, for the system in the
hydrodynamic regime $\omega\tau < 1$, the structure factor is
exhausted by the hydrodynamic sound mode (first sound) even above
the superfluid transition. Therefore, the identification of the
superfluid transition from the density response requires to be in the
collisionless regime, which is reached for  sufficiently weak
interactions.

Within recent experiments \cite{greiner03,grimm04}, the molecular
binding energy in  the BEC limit was determined by a r.f. pulse breaking the
molecules and exciting a fermion into a different hyperfine state.
This method has the disadvantage that the interactions between the
fermions in different hyperfine states produces non-trivial energy
shifts. In turn, the structure factors presented here, produce 
excitations within the same hyperfine states. However, the two
hyperfine states are separated by an energy $\hbar \omega_{\uparrow
\downarrow}$ due to Zeeman splitting. Using a driving field
at frequency $\omega_{\uparrow \downarrow}$ with an additional
superimposed modulation frequency $\omega$, i.e., 
$\lambda(t) \sim \cos(\omega_{\uparrow \downarrow}t ) \cos(\omega t)$, then
probes the structure factors $S_{\rs S}(\omega)$ with the number of particles
in each hyperfine state conserved. Such a procedure avoids 
non-trivial energy shifts induced by a change in particle number or
excitation of particles into  different hyperfine state, and  therefore 
represents a suitable setup for a determination of the two particle excitation 
gap $\Delta_{\rs S}$.
The measurement of the structure factors can be
achieved in two different ways: First, the energy transfer $W$ to the
system  satisfies  $W  =\gamma \omega S(\omega,{\bf k})$ with $\gamma$ 
determined by the driving field alone and allows for the determination 
of $S$ from the heating of the system \cite{stamper-kurn99,steinhauer02}. 
For $^6$Li, the characteristic parameters at the Feshbach resonance
are given by $\Delta_{\rs S} \sim E_{\rs F}\sim 1 \mu{\rm K} $ and
$\omega_{\uparrow,\downarrow}\sim 80 {\rm MHz}$, i.e., the structure
factor is accessbile
via Raman transitions or r.f. pulses as and shown in recent
experiments  \cite{greiner03,grimm04,grimmPD}.
Second, the interaction between the driving field and the fermions leads to
the absorbtion and emission of photons with a rate determined  by the structure factor
$S(\omega,{\bf k})$. Therefore, an analysis of the counting statistic
of the probing laser field allows for an {\it in-situ} measurement of
the structure factors.


 {\acknowledgments We thank G.\ Blatter, A.\ Recati, R.\ Grimm and C.\ Chin for
 stimulating discussions. Work at the Universtiy of Innsbruck is
 supported by the Austrian Science Foundation, European Networks, and
 the Institute of Quantum Information.}


\end{document}